\documentclass[a4paper]{amsart}

\usepackage{amssymb,amsmath}
\usepackage{paralist}
\usepackage{graphicx}
	\graphicspath{{./}{figures/PDF/}{figures/EPS/}}
\usepackage{color}
\usepackage{float}

\theoremstyle{plain}
	\newtheorem{theorem}{Theorem}[section]
\theoremstyle{remark}
	\newtheorem{remark}[theorem]{Remark}

\newcommand{\abs}[1]{\left\vert#1\right\vert}
\newcommand{\ave}[1]{\left\langle#1\right\rangle}

\newcommand{\eps}{\varepsilon}
\newcommand{\mc}{\mathcal }
\newcommand{\n}{\mathbf{n}}
\newcommand{\norm}[1]{\left\Vert#1\right\Vert}
\renewcommand{\P}{\mathbb{P}}
\renewcommand{\r}{\mathbb{R}}
\newcommand{\rdue}{\mathbb R^2}
\newcommand{\rtre}{\mathbb R^3}
\newcommand{\rn}{\mathbb R^n}

\begin{document}

\title[(Mathematical) models for Alzheimer's disease]{Microscopic and macroscopic models for the onset and progression of
Alzheimer's disease}

\author{Michiel Bertsch}
\address{Dipartimento di Matematica, Universit\`{a} di Roma ``Tor Vergata'',  Via della Ricerca Scientifica 1, 00133 Rome, Italy, 
and Istituto per le Applicazioni del Calcolo, CNR, Rome}
\email{bertsch@mat.uniroma2.it}

\author{Bruno Franchi}
\address{Dipartimento di Matematica, Universit\`a di Bologna, Piazza di Porta San Donato 5, 40126 Bologna, Italy}
\email{bruno.franchi@unibo.it}

\author{Maria Carla Tesi}
\address{Dipartimento di Matematica, Universit\`a di Bologna, Piazza di Porta San Donato 5, 40126 Bologna, Italy}
\email{mariacarla.tesi@unibo.it}

\author{Andrea Tosin}
\address{Department of Mathematical Sciences ``G. L. Lagrange'', Politecnico di Torino, Corso Duca degli Abruzzi 24, 10129 Turin, Italy}
\email{andrea.tosin@polito.it}

\begin{abstract}

In the first part of
this paper we review a mathematical model 
for the onset and progression of Alzheimer's disease (AD) that  
was developed in subsequent steps over several years. The
model is meant to describe the evolution of AD \emph{in vivo}. 
In \cite{AFMT} we treated
the problem at a microscopic scale,
where the typical length scale is
a multiple of the size of the soma of a single neuron. 
Subsequently, in~\cite{BFMTT_MMB} we concentrated on the
macroscopic scale, where 
brain neurons are regarded as a continuous medium,  structured  by their degree of malfunctioning. 

In the second part of the paper we consider the relation between the microscopic and the macroscopic models.
In particular we show under which assumptions the kinetic transport equation, which in the macroscopic model governs 
the evolution of the probability measure for the degree of malfunctioning of neurons, can be derived from a 
particle-based  setting. 

The models
are based on aggregation and diffusion equations for $\beta$ Amyloid, a protein fragment that 
healthy brains regularly produce and eliminate. 
In case of dementia A$\beta$ monomers are no longer properly washed out and begin to coalesce forming eventually plaques. 
Two different mechanisms are assumed to be relevant for the temporal evolution of the disease:
\begin{inparaenum}[i)]
\item diffusion and agglomeration of soluble polymers of amyloid, produced by damaged neurons;
\item neuron-to-neuron prion-like transmission.
\end{inparaenum}

In the microscopic model we consider basically  
mechanism i), modelling it by a system of Smoluchowski equations for the amyloid 
concentration (describing the agglomeration phenomenon), with the addition of a diffusion term as well as of a source term on the 
neuronal membrane.
At the macroscopic level instead we model
processes i) and ii) by a system of Smoluchowski equations for the amyloid concentration, 
coupled to a kinetic-type transport equation for the distribution function of the degree of malfunctioning of the neurons. The second 
equation contains an integral term describing the random onset of the disease as a jump process localized in particularly sensitive areas 
of the brain. 

Even though we deliberately neglected many aspects of the complexity of the brain and 
the disease, numerical simulations are
in both cases (microscopic and macroscopic) in good qualitative agreement with clinical data.
\end{abstract}

\maketitle

\section{Introduction}
The aim of the present paper is 
twofold:
to provide an overview of the research carried on in the last few years by several authors in different 
and variated collaborations on both microscopic and macroscopic
mathematical models for Alzheimer's
disease (AD) in the human brain~\cite{AFMT,FT_torino,BFMTT_MMB,FL,FL_wheeden}, 
and to present a new result about the consistency of the microscopic and the macroscopic model. AD
has a huge social and economic impact~
\cite{hurd_et_al,Blennow_et_al,mattson}. 
Until 2040 its global prevalence, estimated as high as 
44 millions in 2015,
is expected to double every 20 years~\cite{reitz_etal_epidemiology}. 
Not by chance AD-related issues belong to the cutting edge of scientific research. Apart from the classical \emph{in vivo} and \emph{in vitro} 
approaches, there is increasing interest in \emph{in silico} research, based on mathematical modelling and computer simulations~\cite{Urbanc19991330,Cruz08071997,good_murphy,Murphy_Pallitto,EDELSTEINKESHET2002301,HHPW,HF}.

To cover the diverse facets of the AD in a single model, different spatial and temporal scales must be taken into account: microscopic spatial scales to describe the role of the neurons, macroscopic spatial and short temporal (minutes, hours) scales for the description of  the relevant diffusion processes in the brain, and large temporal scales (years, decades) for the description of the global development of AD. In~\cite{AFMT,FT_torino} the authors began by attacking the problem at a microscopic scale, that is by considering as size scale of the model a multiple of the size of the soma of a single neuron (from 4 to 100 $\mu$m). Subsequently, see~\cite{BFMTT_MMB}, the authors concentrated on a macroscopic scale, where 
{ they treat}  
brain neurons as a continuous medium, and structure them by their degree of malfunctioning. Mathematically, the bridge between the two models is provided using two quite different techniques:  through the so-called homogenisation technique  in~\cite{FL,FL_wheeden}; and by adapting some arguments of the modern Boltzmann-type kinetic theory for multi-agent systems in Section ~\ref{bridge} of this paper.

Following closely the biomedical literature on the AD, 
 we 
briefly describe the processes (both microscopic and macroscopic) which we 
include in our models. 

In the neurons and their interconnections several \emph{microscopic phenomena} take place. It is largely accepted that beta amyloid (A$\beta$), especially its  highly toxic oligomeric isoforms A$\beta_{40}$ and A$\beta_{42}$, play an important role in the process of the cerebral damage (the so-called \emph{amyloid cascade hypothesis}~\cite{karran_et_al}). In our papers
we focus on  the role of A$\beta_{42}$  in its \emph{soluble} form, which recently has been suggested to be the principal cause of neuronal death and eventually dementia~\cite{walsh}.
Indeed nowadays  there are several evidences, such as enzyme-linked-immunosorbent assays (ELISAs) and mass spectrometry analysis, suggesting that the presence of
plaques is not related to the severity of the AD. On the other hand, high levels of soluble A$\beta$ correlate much better with the presence and degree of cognitive deficits than plaque statistics. 
As a matter of fact some authors (see for instance~\cite{haass_delkoe}) overturn the traditional perspective, claiming that large aggregates of A$\beta$ can actually be inert or even protective to
healthy neurons.

At the level of the neuronal membrane, monomeric A$\beta$ peptides originate from the proteolytic cleavage of a transmembrane glycoprotein, the amyloid precursor protein (APP). By unknown and partially genetic reasons, some neurons present an unbalance between produced and cleared A$\beta$ (we refer to such neurons as damaged neurons). In addition, it has been proposed that neuronal damage spreads in the neuronal net through a neuron-to-neuron prion-like propagation mechanism~\cite{Braak_DelTredici,raj_kuceyeski_weiner}.

On the other hand, \emph{macroscopic phenomena} take place at the level of the cerebral tissue. The monomeric A$\beta$ produced by damaged neurons diffuses through the microscopic tortuosity of the brain tissue and undergoes a process of agglomeration, leading eventually to the formation of  long, insoluble amyloid fibrils, which accumulate in spherical 
deposits known as senile plaques. In addition, soluble A$\beta$ shows a multiple neurotoxic effect: it  induces a general inflammation that activates the microglia which in turn secretes proinflammatory innate cytokines~\cite{griffin_etal_cytokine} and, at the same time, increases intracellular calcium levels~\cite{good_murphy} yielding ultimately apoptosis and neuronal death.

The mathematical models which we derive in Sections~\ref{micro} and~\ref{macro} do not describe all the above-mentioned phenomena involved in the pathological process of the AD. They also neglect
as well other additional phenomena, that we do not even mention. For example, we do not enter  the details of the tortuosity of the brain tissue, we neglect the action of the $\tau$-protein, we simplify the role of the microglia, and neglect its multifaceted mechanism. In fact, we simply assume that high levels of soluble amyloid are toxic for neurons at all scales. Our primary goal was to overcome the fundamental mathematical difficulties and set the basis for a highly flexible 
model,
which can be easily fine-tuned to include other issues. On the other hand,
when we work at macroscopic scale we take into account also a neuron-to-neuron prion-like propagation mechanism
(\cite{raj_kuceyeski_weiner,Braak_DelTredici}),

The models are minimal but effective: the numerical simulations produce \emph{a posteriori} images and graphs which are in good qualitative agreement with clinical findings and confirm the validity of our assumptions. They also capture, at different scales, the cerebral damage in the early stage of the Mild Cognitive Impairment (MCI~\cite{petersen_et_al}). 

In Section~\ref{bridge} we derive the equation for the progression of AD in the macroscopic model from a microscopic description of three main biophysical processes, among those recalled above. Namely, a prion-like spread of the disease over the neural network, the poisoning effect of soluble A$\beta$ polymers diffusing in the brain tissue and stochastic jumps in the level of neuron malfunctioning due to uncontrolled causes, such as e.g. external agents or genetic factors. We use mathematical techniques coming from the modern Boltzmann-type kinetic theory for multi-agent systems~\cite{pareschi2013BOOK}, such as microscopic binary interaction schemes and mean-field asymptotic limits.

In Section~\ref{discussion} we highlight some shortage of the present approach and we discuss possible extensions of the models, 
inspired by future research directions.

\medskip

\section{Mathematical model at the microscopic scale}
\label{micro}
When aiming at producing  mathematical models of biological phenomena we have to fix preliminarily a spatial scale, as well as a time scale. Thus, we consider a portion of the
cerebral cortex comparable in size to the size of a neuron, and we omit both the description of intracellular phenomena and clinical manifestations of the disease at a macroscopic scale, which will be considered instead in the model at the macroscopic scale. On the other hand, the experimental data of~\cite{Meyer-Luhmann_thesis,Meyer-Luhmann_nature} show
that the process of plaques formation takes few days and therefore our temporal scale is chosen of the order of hours. In particular, no anatomical alteration of the neurons and of the surrounding cerebral tissue is taken into account.

The portion of cerebral tissue we consider is represented by a bounded smooth region $\Omega_0\subset\rtre$ (or $\Omega_0\subset\rdue$ in numerical simulations to reduce the computational complexity). To fix our ideas, we can think that the diameter of $\Omega_0$ is of the order of $10$ $\mu$m. The neurons are represented by a family of regular regions $\Omega_j$ such that
\begin{enumerate}
\item $\overline{\Omega}_j\subset \Omega_0$ if $ j=1,\dots, M$;
\item $\overline{\Omega}_i\cap  \overline{\Omega}_j=\emptyset$ if $i\ne j$.
\end{enumerate}
We set
$$ \Omega:=\Omega_0\setminus \bigcup_{j=1}^M \overline{\Omega}_i. $$

To describe the evolution of the amyloid in $\Omega$, we consider a vector-valued function $u=(u_1,\dots,u_N)$, where $N\in\mathbb N$, $u_m=u_m(\tau,x)$, $m=1,\dots,N$, $x\in\Omega$ is the space variable and $\tau\ge 0$ is the time variable. If $1\le m<N-1$ then $u_m(\tau,x)$ denotes the (molar) concentration at time $\tau\ge 0$ and point $x\in\Omega$ of A$\beta$ assemblies of polymers of length $m$. In addition, $u_N$ takes into account aggregations of more than $N-1$ monomers. Although $u_N$ has a different meaning from the other $u_m$'s, we keep the same letter $u$ in order to avoid cumbersome notations. Clusters of oligomers of length $\geq N$ (fibrils) may be thought of as a medical parameter (the plaques), clinically observable through PIB-PET (Pittsburgh compound B - PET~\cite{Nordberg}).

Coherently with this choice of the scales, it is coherent to assume that the diffusion of A$\beta$ in $\Omega$ is uniform, and therefore employ the usual Fourier linear diffusion equation
(see, for instance,~\cite{Nicholson1998207}).

In addition, we describe the agglomeration phenomena by means of the so-called finite Smoluchowski system of equations with diffusion. Classical references are~\cite{smoluchowski,drake}; applications of Smoluchowski system to the description of the agglomeration of A$\beta$ amyloid appeared for the first time in~\cite{Murphy_Pallitto}.

The production of A$\beta$ in the monomeric form at the level of neuron membranes is modelled by a an inhomogeneous Neumann condition on $\partial\Omega_j$, the boundary of $\Omega_j$, for $j=1,\dots,M$. Finally, a homogeneous Neumann condition on $\partial\Omega_0$ is meant to neglect the neighbouring cerebral regions.

Thus, we are led to the following Cauchy-Neumann problem

\begin{equation} \label{eq m=1}
	\left\{
	\begin{array}{c}
		\partial_{ \tau}u_1=d_1\Delta_xu_1-u_1\sum_{j=1}^{N}a_{1,j}u_j \\
		\partial_\nu u_1=\psi_0\equiv0 \quad \mbox{on } \partial\Omega_0 \\
		\partial_\nu u_1=\psi_j \quad \mbox{on } \partial\Omega_j, j=1,\dots,M \\
		u_1(x,0)=U_1(x)\ge 0,
	\end{array}
	\right.
\end{equation}
where $0\le\psi_j\le 1$ is a smooth function for $j=1,\dots, M$ describing the production of the amyloid near the membrane of the neuron.

We only take into account neurons affected by the disease, i.e. we assume $\psi_j\not\equiv 0$ for $j=1,\dots,M$. Moreover, to avoid technicalities, we assume that $U_1$ is smooth, more precisely $U_1\in\mathbf C^{2+\alpha}(\bar\Omega)$ for some $\alpha\in (0,1)$, and that $\partial_\nu U_1
=\psi_j$ on $\partial{\Omega_j}$, $j=0,\dots,M$.

In addition, if $1<m<N$,
\begin{equation} \label{eq m>1}
	\left\{
	\begin{array}{c}
		\partial_\tau u_m=d_m\Delta_xu_m-u_m\sum_{j=1}^{N}a_{m,j}u_j+\frac12\sum_{j=1}^{m-1}a_{j,m-j}u_ju_{m-j} \\
		\partial_\nu u_m =0 \quad\mbox{on } \partial\Omega_0 \\
		\partial_\nu u_m=0 \quad\mbox{on } \partial\Omega_j, j=1,\dots, M \\
		u_m(x,0)=0,
	 \end{array}
	\right.
\end{equation}
and
\begin{equation} \label{eq m=N}
	\left\{
	\begin{array}{c}
		\partial_\tau u_N=d_N\Delta_xu_N+\frac12\sum_{j+k\ge N, k<N,j<N}a_{j,k}u_ju_{k} \\
		\partial_\nu u_N=0 \quad\mbox{on } \partial\Omega_0 \\
		\partial_\nu u_N=0 \quad\mbox{on } \partial\Omega_j, j=1,\dots, M \\
		u_N(x,0)=0,
	\end{array}
	\right.
\end{equation}
where $d_j>0$, $j=1,\dots,N$ and $a_{i,j}=a_{j,i}>0$, $i,j=1,\dots,N$ (but $a_{N,N}=0$).

We assume that the diffusion coefficients $d_j$ are small when $j$ is large, since big assemblies do not move. In fact, the diffusion coefficient of a soluble peptide scales approximately as the reciprocal of the cube root of its molecular weight (see~\cite{Goodhill1997,Nicholson1998207}).

The form of the coefficients $a_{i,j}$ (the coagulation rates) considered in~\cite[formula (13)]{Murphy_Pallitto} rely on sophisticated statistical mechanics considerations (see also~\cite{hill,tomsky_murphy}). In our numerical simulations, we use a slightly approximate form of these coefficients, taking $a_{i,j}=\dfrac{\alpha}{ij}$ where $\alpha>0$. In fact, this approximation basically consists in neglecting logarithmic terms in front of linear ones for large $i,j$. Concerning the coefficient $a_{N,N}$, it is clearly consistent with experimental data to assume $a_{N,N}=0$ for large $N$, which is equivalent to say that large oligomers do not aggregate with each other.

In our simulations we identify senile plaques with the sets $\{x\in\Omega\,:\,u_N(\tau,x)>c>0\}$. The following picture is provided by a numerical implementation of the model ~\eqref{eq m=1},~\eqref{eq m>1},~\eqref{eq m=N}. As in clinical observations, plaques grow near a neuron (the circle in the picture). The picture has been obtained by taking appropriate level sets of $u_N(\tau,\cdot)$.
\begin{figure}[H]
\centering
\includegraphics[height=6.5cm,width=8.5cm]{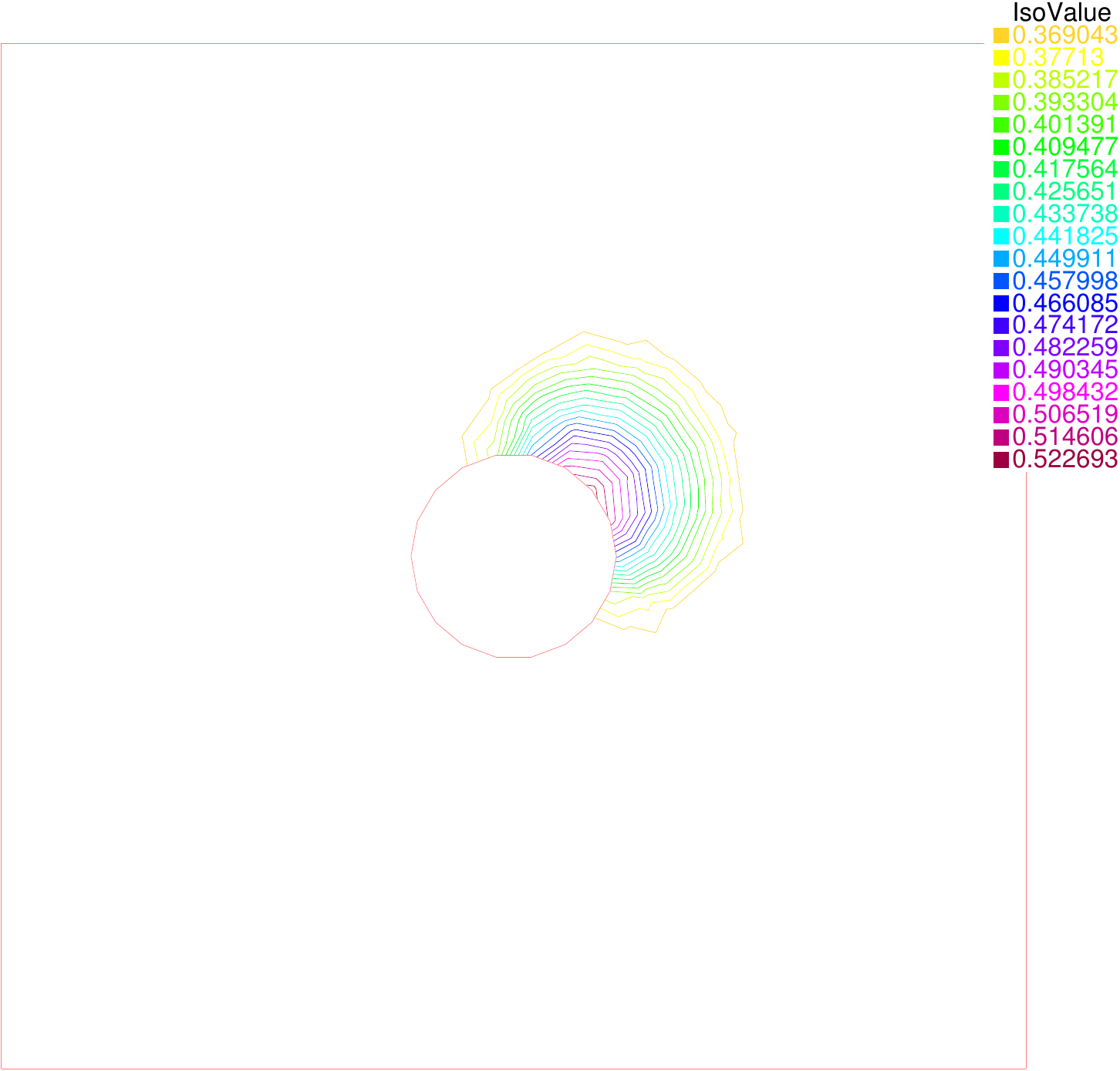}
\caption{Plaque generated with $N=16$, $\alpha=10$, $U_1\equiv 0$, $\psi=0.5$.}
\label{fig:1}
\end{figure}

We notice that our model leads to a smooth shape of the senile plaques (because of standard regularity properties of diffusion equations), in disagreement with evidences found \emph{in vivo}. This may be explained by Figure 3 in~\cite{EDELSTEINKESHET2002301} and related comments on the role of the microglia.

Besides numerical simulations, the main result obtained in~\cite{AFMT} for this model is the following existence theorem:

\begin{theorem} \label{existence}
For all $T>0$  the Neumann-Cauchy problem~\eqref{eq m=1},~\eqref{eq m>1},~\eqref{eq m=N} has a unique classical positive solution $u\in\mathbf C^{1+\alpha/2, 2+\alpha}([0,T]\times \bar{\Omega})$.
\end{theorem}

\bigskip

\section{Mathematical model at the macroscopic scale}
\label{macro}
We identify now a large portion of the cerebral tissue with a 3-dimensional region $\Omega$, with diameter of $\Omega$ of the order of $10$ cm. As for the time scale, a new phenomenon occurs: two temporal scales are needed to simulate the evolution of the disease over a period of years, i.e.  besides the short (i.e., rapid) $\tau$-scale (whose unit time coincides with hours) for the diffusion and agglomeration of A$\beta$~\cite{Meyer-Luhmann_nature} that we used for the microscopic model, we need a long (i.e., slow) $t$-scale (whose unit time coincides with several months) to take into account the progression of AD. We can write the relation between the two scales as $\Delta{t}=\eps\Delta\tau$ for a small parameter $\eps\ll 1$.

At the macroscopic scale, the boundary vale problem for monomers~\eqref{eq m=1} must have a different form. Indeed, the information given on the microscale by the non-homogeneous
Neumann boundary condition is transferred into a source term $\mathcal F$ appearing in the macroscopic equation. This is due to the fact that at this scale neurons are reduced to points. Therefore, we have the following macroscopic equation for monomers:
\begin{equation} \label{amyloid equation: m=1}
	\partial_\tau u_1=d_1\Delta_x u_1-u_1\sum_{j=1}^{N}a_{1,j}u_j+\mathcal{F}
\end{equation}
while the equations in~\eqref{eq m>1} and~\eqref{eq m=N} remain unchanged.

Mathematically, the transition from system~\eqref{eq m=1} to equation~\eqref{amyloid equation: m=1} has been obtained by a two-scale homogenisation procedure described in~\cite{FL} and~\cite{FL_wheeden}.

The source term $\mathcal{F}$ in~\eqref{amyloid equation: m=1} will depend on the health state of the neurons. We model the \emph{degree of malfunctioning} of a neuron with a parameter $a$ ranging from $0$ to $1$: $a$ close to $0$ stands for ``the neuron is healthy'' whereas $a$ close to $1$ for ``the neuron is dead''. This parameter, although introduced for the sake of mathematical modelling (see also~\cite{raj_kuceyeski_weiner}), can be compared with medical images from Fluorodeoxyglucose PET (FDG-PET
\cite{mosconi_et_al}).

For fixed $x\in\Omega$ and $t\geq 0$, let $f(x,\,a,\,t)$ be a probability measure, supported in $[0,1]$, that indicates the \emph{fraction} of neurons close to $x$ with degree of malfunctioning  between $a$ and $a+da$ at time $t$. From now on, we denote by $X_{[0,1]}$ the space of probability measures on $\mathbb R$ that are supported in $[0,1]$.

Since A$\beta$ monomers are produced by neurons and the production increases if neurons are damaged, we choose in~\eqref{amyloid equation: m=1}
\begin{equation}\label{mathcal F}
	\mathcal{F}=\mathcal{F}[f]=C_{\mathcal F}\int_0^1(\mu_0+a)(1-a)df(x,\,a,\,t).
\end{equation}
The small constant $\mu_0>0$ accounts for the physiologic A$\beta$ production by healthy neurons, and the factor $1-a$
for the fact that dead neurons do not produce amyloid. 

The progression of AD occurs in the slow time scale $t$, over decades, and is determined by the \emph{deterioration rate} $v=v(x,\,a,\,t)$ of the health state of the neurons
through the continuity equation:
\begin{equation}\label{health macro}
	\partial_t f+\partial_a(fv[f])=0.
\end{equation}
Here $v[f]$ indicates that the deterioration rate depends on $f$ itself. 

We assume that 
\begin{equation}\label{velocity bis}
	v[f]=\int_0^1\mathcal{G}(x,\,a,\,b)\, df(x,\,b,\,t)+\mathcal{S}(x,\,a,\,u_1(x,\,\tau),\,\dots,\,u_{N-1}(x,\,\tau)).
\end{equation}
The notation $\mathcal G$ takes into account the spreading of the disease by proximity, while $\mathcal{S}$  models the action of toxic A$\beta$ oligomers, ultimately leading to apoptosis. For instance, we can choose
\begin{equation}\label{mathcal G}
	\mathcal{G}(x,\,a,\,b)=C_{\mathcal G}(b-a)^{+},
\end{equation}
and
\begin{equation}\label{mathcal S}
	\mathcal{S}=C_{\mathcal S}(1-a){\left(\sum\limits_{m=1}^{N-1}mu_m(x,\,\tau)-\overline{U}\right)}^{+}.
\end{equation}
The threshold $\overline{U}>0$ indicates the minimal amount of toxic A$\beta$ needed to damage neurons, assuming that the toxicity of soluble $A\beta$-polymers does not depend on $m$. In reality length dependence has been observed~\cite{Ono_et_al}, but, to our best knowledge, quantitative data are only available for very short molecules (see~\cite[Table 2]{Ono_et_al}). For long molecules any analytic expression would be arbitrary.

At this point, we stress that equation~\eqref{health macro}, by its own nature, fails to describe the onset of the disease. To describe the onset of AD we assume that in small, randomly chosen parts of the cerebral tissue, concentrated for instance in the hippocampus, the degree of malfunctioning of neurons randomly jumps to higher values due to external agents or genetic factors. This leads to an additional term in the equation for $f$,
\begin{equation*}
	\partial_t f+\partial_a\left(fv[f]\right)=J[f],
\end{equation*}
where
\begin{equation} \label{J formula}
	J[f]=\eta\left(\int_0^1 P(t,\,a_\ast\to a)f(x,\,a_\ast,\,t)\,da_\ast-f(x,\,a,\,t)\right)\chi(x,\,t).
\end{equation}
Here, $P(t,\,a_\ast\to a)$ is the probability to jump from the state $a_\ast$ to a state $a\in[0,\,1]$ (obviously, $P(t,\,a_\ast\to a)=0$ if $a<a_\ast$), $\chi(x,\,t)$ describes the random jump distribution, and $\eta$ is the jump frequency. For instance we can choose 
\begin{equation}\label{P}
	P(t,\,a_\ast\to a)\equiv P(a_\ast\to a)=
		\begin{cases}
			\dfrac{2}{1-a_\ast} & \text{if\ } a_\ast\leq a\leq\dfrac{1+a_\ast}{2} \\
			0 & \text{otherwise},
		\end{cases}
\end{equation}
and neglect randomness, taking $\chi(x,\,t)\equiv\chi(x)$, concentrated in the hippocampus.

Finally, to model the phagocytic activity of the microglia as well as other bulk clearance processes~\cite{Iliff_et_al}, we add a term $-\sigma_mu_m$
in equations  \eqref{eq m>1}, \eqref{eq m=N} and \eqref{amyloid equation: m=1}, where $\sigma_m>0$.

We consider a transversal section (i.e. a horizontal planar section) of the brain that can be compared with radiological imaging (see, e.g.,~\cite[Fig. 1]{BFMTT_MMB}). For the sake of simplicity we schematise the section of the brain as a bounded connected region $\Omega\subset\rdue$, with two inner disjoint ``holes'' representing the sections of the cerebral ventricles. Consistently, we assume that the boundary of $\Omega$ consists of two disjoint parts: an outer boundary $\partial\Omega_\text{out}$ and an inner boundary $\partial\Omega_\text{in}$, i.e. the boundary of the cerebral ventricles  consisting of two disjoint closed simple curves.

Eventually, we are led to the system
\begin{equation} \label{complete system}
	\begin{cases}
		\partial_t f+\partial_a\left(fv[f]\right)=J[f] \\
		\eps\partial_tu_1=d_1\Delta_x u_1-u_1\sum\limits_{j=1}^{N}a_{1,j}u_j+\mathcal{F}[f]-\sigma_1u_1 \\
		\eps\partial_tu_m=d_m\Delta_x u_m+\dfrac{1}{2}\sum\limits_{j=1}^{m-1}a_{j,m-j}u_ju_{m-j} \\
		\phantom{\eps\partial_tu_m=d_m\Delta_x u_m}-u_m\sum\limits_{j=1}^{N}a_{m,j}u_j-\sigma_m u_m & (2\leq m<N) \\
		\eps\partial_tu_N= \dfrac{1}{2}\sum\limits_{\substack{j+k\geq N \\ k,\,j<N}}a_{j,k}u_ju_{k},
	\end{cases}
\end{equation}
with $\tau$ replaced by $\eps^{-1}t$. Since we are interested in longitudinal modelling, we assume that initially, at $t=0$, there is a small uniform distribution of soluble amyloid $u_0=(u_{0,1},\dots,u_{0,N})$.

Thus system~\eqref{complete system} has to be coupled with Cauchy initial data
\begin{equation}\label{IBC}
	\begin{cases}
		\smallskip f(x,a,0)=f_0(x,a)&\text{if }x\in \Omega, \ 0\le a\le 1\\
		\smallskip u_i(x,0)=u_{0,i}(x) &\text{if }x\in \Omega, \ 1\le i\le N,	
		\end{cases}
\end{equation}
where the $u_{0,i}\in C^1(\overline \Omega)$ are nonnegative functions for $i=1,\dots,N$, and $f_0 \in L^\infty(\Omega; X_{[0,1]})$ describes the distribution of the disease at time $t=0$.

On the outer boundary $\partial\Omega_\text{out}$ we assume vanishing normal polymer flow. Therefore we impose a homogeneous Neumann condition for the diffusing amyloid polymers:
\begin{equation}
	-\frac{d_m}{\eps}\Delta_x {u_m}\cdot\n=0 \quad \text{on } \partial\Omega_\text{out}, \quad
		m=1,\,\dots,\,N-1,
	\label{eq:bc_Neumann}
\end{equation}
$\n$ being the outward normal unit vector to $\partial\Omega_\text{out}$. Notice that no boundary condition is required for the concentration $u_N$ of the fibrillar amyloid, since its equation does not feature space dynamics (cf. the last equation in~\eqref{complete system}). On the inner boundary  $\partial\Omega_\text{in}$, that is the boundaries of the cerebral ventricles, we model the removal of A$\beta$ from cerebrospinal fluid (CSF) through the choroid plexus (cf.~\cite{Iliff_et_al,serot_et_al}) by an outward polymer flow proportional to the concentration of the amyloid. For this, we impose a Robin boundary condition of the form:
\begin{equation}
	-\frac{d_m}{\eps}\Delta_x {u_m}\cdot\n=\gamma u_m \quad \text{on } \partial\Omega_\text{in}, \quad
		m=1,\,\dots,\,N-1,
	\label{eq:bc_Robin}
\end{equation}
with $\gamma>0$ a constant.

An existence and uniqueness theorem for system~\eqref{complete system} with Cauchy initial data~\eqref{IBC} and boundary conditions~\eqref{eq:bc_Neumann} and~\eqref{eq:bc_Robin} is proved in~\cite{BFTT_alzheimer}. With our choices of $P$, $\mc G$ and $\mc S$ in~\eqref{P},~\eqref{mathcal G} and~\eqref{mathcal S}, it reads as follows:

\begin{theorem} For all $T>0$ there exist a unique $(N+1)$-ple 
$$ (f,u_1,\cdots,u_N)\in L^{\infty}(\Omega;\,C([0,T];\,X_{[0,1]}))\times C([0,T];\,C^1(\overline{\Omega}))^N, $$
$u_i\ge 0$ for $i=1,\dots,N$, solving~\eqref{complete system} in a weak sense in $[0,T]$, with  Cauchy data~\eqref{IBC} and  boundary data~\eqref{eq:bc_Neumann} and~\eqref{eq:bc_Robin}.
\end{theorem}
In particular, the first equation in~\eqref{complete system} is satisfied in the following weak sense: for a.e. $x\in\Omega$, for $\phi=\phi(x,\cdot,\cdot)\in \mc D(\mathbb R\times [0,T])$ and for all $t\in [0,T]$
$$
	\begin{aligned}
		& \int_0^t\left(\int(\partial_s\phi+v\partial_a\phi)df(x,\cdot,s)+\int \phi dJ(x,\cdot,s)\right)ds \\
		&\qquad =\int \phi(x,\cdot,t)df(x,\cdot,t)-\int \phi (x,\cdot,0)df_0(x,\cdot).
	\end{aligned}
$$

Concerning the outputs of the numerical simulation of~\eqref{complete system} with Cauchy initial data~\eqref{IBC} and boundary conditions~\eqref{eq:bc_Neumann} and~\eqref{eq:bc_Robin}, it is instructive to compare plots of $f$, at different times, with FDG-PET images (see e.g.~\cite{reiman2}): we create a schematic image of a transverse section of the brain and attribute different colors (varying from red to blue) to those parts of the brain where probabilistically the level of malfunctioning lies in different ranges. As in the FDG-PET, the red corresponds to a healthy tissue. Here AD originates only from the hippocampus and propagates, at the beginning, along privileged directions (such as those corresponding to denser neural bundles) mimicked by two triangles. Obviously the details of the numerical output depend on the choice of the constants used in the mathematical model. Performing a considerable amount of numerical runs with different values of the constants in the model, we have reached the conclusion that, at least qualitatively, the behaviour of the solutions does not depend on the precise choice of those constants, as long as their variation is restricted to reasonable ranges. 

\begin{figure}[H]
\centering
\includegraphics[width=5.5cm]{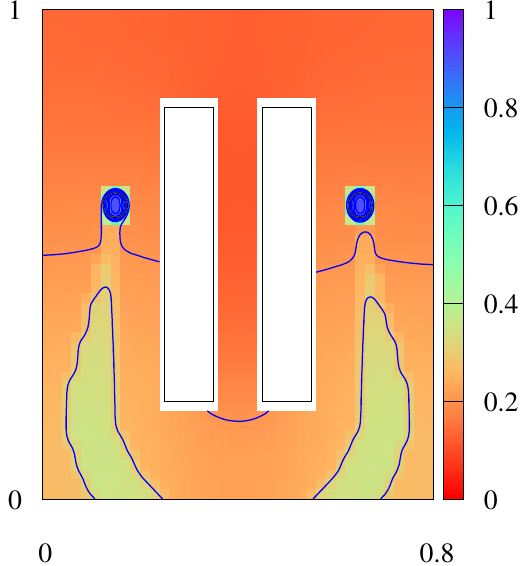}
\qquad
\includegraphics[width=5.5cm]{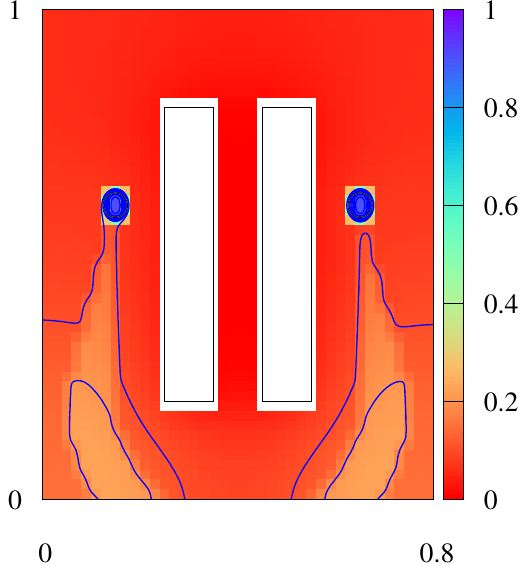}
\caption{Neuron malfunctioning: $\gamma=0.01$ (left), $\gamma=1$ (right), $T=30$.}
\label{fig:diseaseT=30}
\end{figure}

\begin{figure}[H]
\centering
\includegraphics[width=5.5cm]{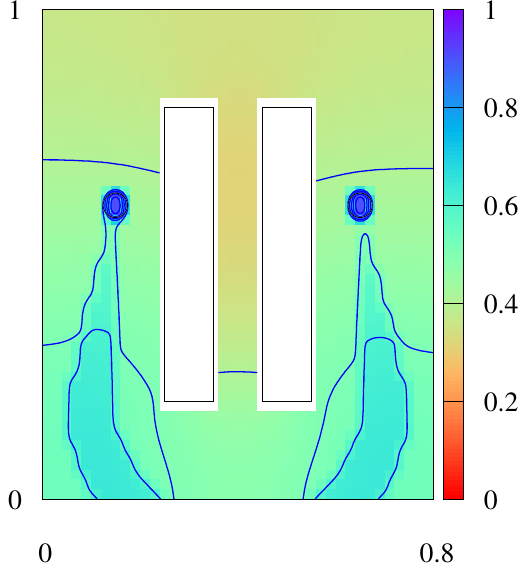}
\qquad
\includegraphics[width=5.5cm]{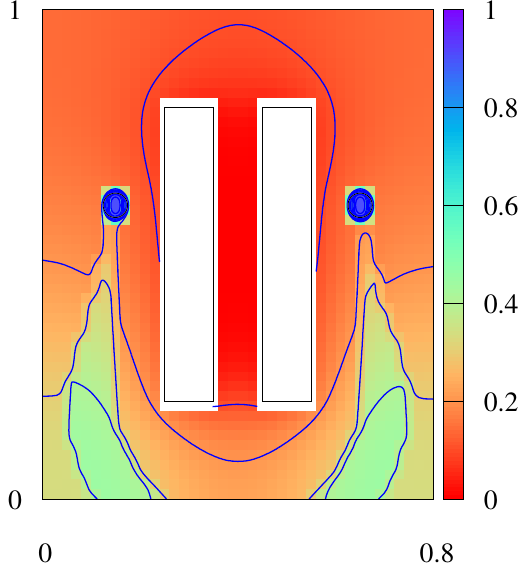}
\caption{Neuron malfunctioning: $\gamma=0.01$ (left), $\gamma=1$ (right), $T=40$.}
\label{fig:diseaseT=40}
\end{figure}

The constant $\gamma$ enters the model through condition~\eqref{eq:bc_Robin} at the boundary of the cerebral ventricles. Smaller values of $\gamma$ mean that less A$\beta$ is removed from the CSF through the choroid plexus. The comparison of the cases $\gamma=1$ and $\gamma=0.01$ becomes quite clear when we create spatial plots of $f$  (taking into account the average degree of malfunctioning of the brain in every point) at fixed computational times $t=T$.   In Figures~\ref{fig:diseaseT=30} and~\ref{fig:diseaseT=40}, where we compare plots of $f$ at, respectively, times $T=30$ and $T=40$ for the two different values of $\gamma =0.01$ and $\gamma=1$, AD originates only from the hippocampus and propagates, at the beginning, along privileged directions (such as those corresponding to denser neural bundles) mimicked by two triangles. 
Two remarks are now in order. First of all, though our images represent a {\sl mean value} of brain activity instead of a single patient's
brain activity, still they show a good agreement with clinical neuroimaging (obviously representing the specific situation of an individual patient). 
See, e.g. \cite{miller2006RR},
reproduced also with permission in \cite{BFMTT_MMB}, Fig. 6. The specificities (both anatomic and physiologic) of the single patient might account for the 
discrepancies between the outputs of our simulations and clinical neuroimaging.
Secondly, we notice that if $\gamma$ becomes smaller (corresponding to a lower rate of
clearance of the amyloid), we observe
a temporal acceleration of the development of the illness; this could indicate the potential importance of the removal of $A\beta$ through the choroid plexus to slow down the temporal development of AD.

\begin{figure}[H]
\centering
\includegraphics[width=5.5cm]{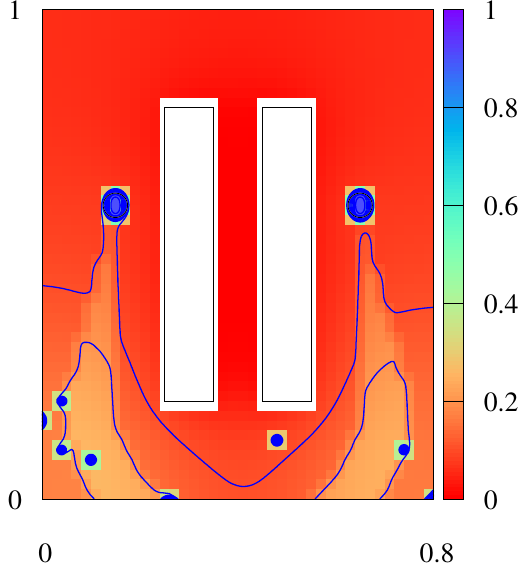}
\qquad
\includegraphics[width=5.5cm]{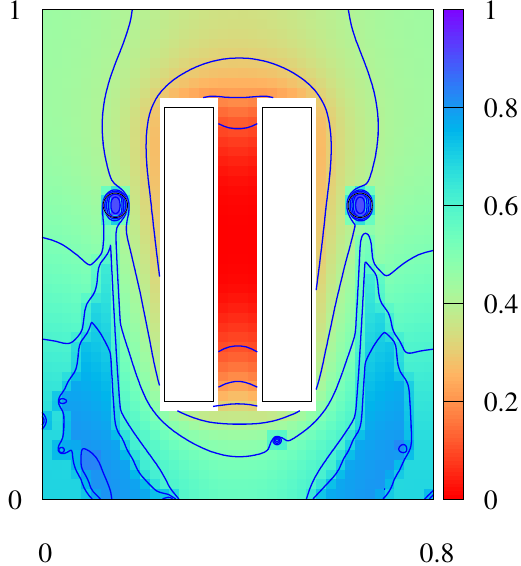}
\caption{Neuron disease with random sources for $\gamma=1$ at $T=30$ (left) and $T=60$ (right).}
\label{fig:random_sources}
\end{figure}

Looking for even more realistic images, we  take now into account the randomness of the spatial distribution of the sources of the disease. Therefore we perform some runs where the AD does not only originate from the hippocampus, but also from several sources of A$\beta$ randomly distributed in the occipital part of the brain. We report the outputs of such runs, for $\gamma=1$ and two different values of time $T$, in Figure~\ref{fig:random_sources}. The randomly distributed sources appear as  the small blue spots.

Comparing the random sources case with the one in which AD originates only in  the hippocampus, for the same values of  $\gamma=1$ and $T=30$, it is clear that the brain sickness is more advanced when the number of sources is increased.

\bigskip

\section{Derivation of the macroscopic equation for the progression of the disease}
\label{bridge}
While the macroscopic system of Smoluchowski equations in~\eqref{complete system} has been obtained from a smaller neuron-size scale through the homogenisation technique described in~\cite{FL,FL_wheeden}, the mesoscopic equation for the distribution $f$ of the disease has been so far postulated on a mainly heuristic basis. In this section we provide its derivation  from more fundamental particle-based dynamics, by adapting some arguments of the modern Boltzmann-type kinetic theory for multi-agent systems~\cite{pareschi2013BOOK}.

\subsection{Particle-based neuron dynamics}
\label{sect:micro.dyn}
Let $\tau\geq 0$ be the short (i.e. rapid) time variable, like in Section~\ref{macro}, and $\Omega\subset\rn$ ($n=2,\,3$) a bounded subset of the physical space representing the brain, or possibly a two-dimensional section of it. We denote by $X\in\Omega$ the position of a neuron in the brain and by $A_\tau\in [0,\,1]$ its degree of malfunctioning at time $\tau$, which we assume to evolve according to the main biophysical mechanisms mentioned in the Introduction:
\begin{itemize}
\item a \emph{neuron-to-neuron prion-like transmission} of the disease regarded as a binary interaction with a neighbouring neuron in the point $Y\in\Omega$ with degree of malfunctioning $B_\tau\in [0,\,1]$ at time $\tau$. We model the effect of such a binary interaction by a term of the form
$$ H_{X,Y}\mathcal{G}(X,\,A_\tau,\,B_\tau), $$
where $\mathcal{G}:\Omega\times [0,\,1]\times [0,\,1]\to [0,\,1]$ is a prescribed function which accounts for the prionic transmission of the disease and $H_{X,Y}\in\{0,\,1\}$ is a variable which describes the structure of the neural network. Specifically, $H_{X,Y}=1$ if the neurons in $X$ and $Y$ are connected by a synapse while $H_{X,Y}=0$ if they are not.

Due to the extremely complicated structure of the neural network, we consider a simple probabilistic description of it by assuming that $H_{X,Y}$ is a Bernoulli random variable parameterised by $X$, $Y$, in the sense that its law is
\begin{equation}
	\P(H_{X,Y}=1)=h(X,\,Y), \qquad \P(H_{X,Y}=0)=1-h(X,\,Y)
	\label{eq:Prob_H}
\end{equation}
for a given $h:\Omega\times\Omega\to [0,\,1]$ such that $h(x,\,y)=h(y,\,x)$ for all $x,\,y\in\Omega$.

 On the other hand, as an approximation we explicitly disregard the variability in time of the connections among the neurons;
\item the \emph{poisoning effect of soluble A$\beta$ polymers} diffusing in the brain tissue, which we model by a function
$$ \mathcal{S}=\mathcal{S}(X,\,A_\tau,\,u(X,\,\tau)), $$
where we have set $u(x,\,\tau):=(u_1(x,\,\tau),\,\dots,\,u_{N-1}(x,\,\tau))$, $x\in\Omega$, for brevity;
\item \emph{stochastic jumps} in the degree of malfunctioning due to uncontrolled causes, such as external agents or genetic factors, which we model by means of a random variable $J_\tau$ such that
$$ 0\leq J_\tau\leq 1-A_\tau, $$
as in no case the new degree of malfunctioning after the jump, i.e. $A_\tau+J_\tau$, can be greater than $1$.
\end{itemize}

In order to introduce a rule for the time variation of $A_\tau$ we assume that in a short time interval $\Delta{\tau}>0$ there is a probability proportional to $\Delta{\tau}$ that the neuron undergoes any of the mechanisms mentioned above. Furthermore, we assume that each mechanism is independent of the others. A simple way to formalise this is to introduce three independent Bernoulli random variables $T_\nu,\,T_\mu,\,T_\eta\in\{0,\,1\}$ such that
$$ \P(T_\kappa=1)=\kappa\Delta{\tau}, \quad \P(T_\kappa=0)=1-\kappa\Delta{\tau}, \qquad \kappa=\nu,\,\mu,\,\eta, $$
where $\nu,\,\mu,\,\eta>0$ are the frequencies associated to each of the mechanisms above while the time interval has to be chosen in such a way that $\Delta{\tau}<1/\max\{\nu,\,\mu,\,\eta\}$. Under this assumption we set
\begin{equation}
	A_{\tau+\Delta{\tau}}=A_\tau+T_\nu H_{X,Y}\mathcal{G}(X,\,A_\tau,\,B_\tau)+T_\mu\mathcal{S}(X,\,A_\tau,\,u(X,\,\tau))
		+T_\eta J_\tau.
	\label{eq:micro_stochastic}
\end{equation}

\subsection{Boltzmann-type kinetic description}
\label{sect:boltzmann}
Let $\varphi=\varphi(x,\,a):\Omega\times [0,\,1]\to\r$ be a test function representing any observable quantity that can be computed out of the microscopic state $(X,\,A_\tau)$ of a neuron. From~\eqref{eq:micro_stochastic} we get:
$$ \varphi(X,\,A_{\tau+\Delta{\tau}})=\varphi(X,\,A_\tau+T_\nu H_{X,Y}\mathcal{G}(X,\,A_\tau,\,B_\tau)
		+T_\mu\mathcal{S}(X,\,A_\tau,\,u(X,\,\tau))+T_\eta J_\tau), $$
whence, averaging both sides and computing first the mean with respect to the variables $T_\nu,\,T_\mu,\,T_\eta$,
\begin{align}
	\begin{aligned}[t]
		\ave{\varphi(X,\,A_{\tau+\Delta{\tau}})}=\ave{\varphi(X,\,A_\tau)}
			+\Delta{\tau}\Bigl[&\nu\ave{\varphi(X,\,A_\tau+H_{X,Y}\mathcal{G}(X,\,A_\tau,\,B_\tau))} \\
		& +\mu\ave{\varphi(X,\,A_\tau+\mathcal{S}(X,\,A_\tau,\,u(X,\,\tau)))} \\
		& +\eta\ave{\varphi(X,\,A_\tau+J_\tau)} \\
		& -(\nu+\mu+\eta)\ave{\varphi(X,\,A_\tau)}\Bigr]+o(\Delta{\tau}),
	\end{aligned}
	\label{eq:Boltz.first_ave}
\end{align}
where $\ave{\cdot}$ denotes the average. Furthermore, using~\eqref{eq:Prob_H} to compute the mean with respect to $H_{X,Y}$ in the first term in brackets at the right-hand side, we obtain
\begin{align*}
	\langle\varphi(X,\,A_\tau+H_{X,Y} &\mathcal{G}(X,\,A_\tau,\,B_\tau))\rangle \\
	&= \ave{\varphi(X,\,A_\tau+\mathcal{G}(X,\,A_\tau,\,B_\tau))h(X,\,Y)} \\
	&\phantom{=} +\ave{\varphi(X,\,A_\tau)(1-h(X,\,Y))},
\end{align*}
hence~\eqref{eq:Boltz.first_ave} specialises in
\begin{align*}
	& \frac{\ave{\varphi(X,\,A_{\tau+\Delta{\tau}})}-\ave{\varphi(X,\,A_\tau)}}{\Delta{\tau}} \\
	&\hspace{24mm} =\nu\ave{\bigl(\varphi(X,\,A_\tau+\mathcal{G}(X,\,A_\tau,\,B_\tau))-\varphi(X,\,A_\tau)\bigr)h(X,\,Y)} \\
	&\hspace{24mm}\phantom{=} +\mu\ave{\varphi(X,\,A_\tau+\mathcal{S}(X,\,A_\tau,\,u(X,\,\tau)))} \\
	&\hspace{24mm}\phantom{=} +\eta\ave{\varphi(X,\,A_\tau+J_\tau)}-(\mu+\eta)\ave{\varphi(X,\,A_\tau)}+o(1).
\end{align*}
In the limit $\Delta{\tau}\to 0^+$ this produces the continuous-in-time master equation
\begin{align}
	\begin{aligned}[t]
		\frac{d}{d\tau}\ave{\varphi(X,\,A_\tau)} &= \nu\ave{\bigl(\varphi(X,\,A_\tau+\mathcal{G}(X,\,A_\tau,\,B_\tau))-\varphi(X,\,A_\tau)\bigr)h(X,\,Y)} \\
		&\phantom{=} +\mu\ave{\varphi(X,\,A_\tau+\mathcal{S}(X,\,A_\tau,\,u(X,\,\tau)))-\varphi(X,\,A_\tau)} \\
		&\phantom{=} +\eta\ave{\varphi(X,\,A_\tau+J_\tau)-\varphi(X,\,A_\tau)}.
	\end{aligned}
	\label{eq:micro_averaged}
\end{align}
	
Let us now introduce the probability density function
$$ g=g(x,\,a,\,\tau):\Omega\times [0,\,1]\times\r_+\to\r_+ $$
of the microscopic state $(X,\,A_\tau)$, i.e. $g(x,\,a,\,\tau)\,dx\,da$ is the fraction of neurons which at time $\tau$ are in the infinitesimal volume $dx$ centred at $x\in\Omega$ with a degree of malfunctioning in $[a,\,a+da]$. In the spirit of a \emph{Boltzmann-type ansatz}, we assume that the processes $(X,\,A_\tau)$ and $(Y,\,B_\tau)$ are independent, so that their joint law is $g(x,\,a,\,\tau)g(y,\,b,\,\tau)$, cf. the next Remark~\ref{rem:Boltz.ansatz}. Moreover, we denote by $p(\tau,\,j\vert x,\,a)$, $0\leq j\leq 1-a$, the law of $J_\tau$ conditioned to $(X,\,A_\tau)$, which is such that
\begin{equation}
	\int_0^{1-a}p(\tau,\,j\vert x,\,a)\,dj=1, \qquad \forall\,x\in\Omega,\ a\in [0,\,1],\ \tau\geq 0.
	\label{eq:int.p}
\end{equation}
In view of these positions, we compute explicitly the remaining averages in~\eqref{eq:micro_averaged} as:
\begin{align}
	\begin{aligned}[t]
		\frac{d}{d\tau}\int_0^1 &\int_\Omega\varphi(x,\,a)g(x,\,a,\,\tau)\,dx\,da \\
		&= \nu\int_0^1\int_0^1\int_\Omega\int_\Omega(\varphi(x,\,a^\ast)-\varphi(x,\,a))
			h(x,\,y)g(x,\,a,\,\tau)g(y,\,b,\,\tau)\,dx\,dy\,da\,db \\
		&\phantom{=} +\mu\int_0^1\int_\Omega(\varphi(x,\,a^{\ast\ast})-\varphi(x,\,a))g(x,\,a,\,\tau)\,dx\,da \\
		&\phantom{=} +\eta\int_0^1\int_\Omega\int_0^{1-a}(\varphi(x,\,a^{\ast\ast\ast})-\varphi(x,\,a))
			p(\tau,\,j\vert x,\,a)g(x,\,a,\,\tau)\,dj\,dx\,da,
	\end{aligned}
	\label{eq:Boltzmann.g}
\end{align}
where the starred variables denote the state of the neuron after one of the three types of interactions according to~\eqref{eq:micro_averaged}:
\begin{align}
	\begin{aligned}[t]
		a^\ast &= a+\mathcal{G}(x,\,a,\,b) & \text{(prion-like transmission of the disease)} \\
		a^{\ast\ast} &= a+\mathcal{S}(x,\,a,\,u(x,\,\tau)) & \text{(poisoning by $A\beta$ polymers)} \\
		a^{\ast\ast\ast} &= a+j & \text{(stochastic jumps)}.
	\end{aligned}
	\label{eq:int.rules}
\end{align}

Equations~\eqref{eq:Boltzmann.g},~\eqref{eq:int.rules} provide the \emph{Boltzmann-type kinetic description} of the microscopic model formulated in Section~\ref{sect:micro.dyn}.

\begin{remark} \label{rem:Boltz.ansatz}
Inspired by the discussion set forth in~\cite[Chapter 1]{pareschi2013BOOK}, we observe that the assumption of stochastic independence of the states $(X,\,A_\tau)$, $(Y,\,B_\tau)$ is not fully justified from the biological point of view, being mostly dictated by the wish to obtain a closed equation in terms of the sole distribution function $g$. However, as it often happens in this type of problems, one needs to mediate between the high complexity of the biological phenomenon and the possibility to construct a usable, though necessarily approximated, mathematical model. In this respect, the aforesaid assumption should be regarded as a reasonable compromise, which permits a quite complete description and analysis of the evolution of the system.
\end{remark}

\subsection{The quasi-invariant degree of malfunctioning limit}
\label{sect:grazing}
As recalled in Section~\ref{macro}, the progression of AD occurs in a much slower time scale than that of the diffusion and agglomeration of A$\beta$ polymers. This implies that the actual time scale where the macroscopic effects of the progression of AD are observable is much longer than the $\tau$-scale at which the particle dynamics discussed in Sections~\ref{sect:micro.dyn},~\ref{sect:boltzmann} take place. For this reason, as anticipated in Section~\ref{macro} and inspired by the quasi-invariant interaction limits introduced in~\cite{pareschi2013BOOK,toscani2006CMS}, we now define the new time variable
\begin{equation}
	t:=\eps\tau, \qquad 0<\eps\ll 1,
	\label{eq:time.scaling}
\end{equation}
where $\eps$ is a dimensionless parameter. Under this scaling, the typical time of a single particle transition \eqref{eq:int.rules}, which is $O(1)$ in the $\tau$-scale, becomes much shorter in the $t$-scale, precisely $O(\eps)$. Simultaneously, we scale by $\eps$ also the  interactions~\eqref{eq:int.rules}, considering that the effect of a single transition is  attenuated in the longer $t$-scale. In particular, we set
\begin{align}
	\begin{aligned}[t]
		a^\ast &= a+\eps \,\mathcal{G}(x,\,a,\,b) \\
		a^{\ast\ast} &= a+\eps\,\mathcal{S}(x,\,a,\,u(x,\,\tau)). \\
	\end{aligned}
	\label{eq:int.rules.eps}
\end{align}
As far as the stochastic jumps are concerned, we assume instead that the strength of a single jump is independent of the time scale, hence we still have $a^{\ast\ast\ast}=a+j$ also in the $t$-scale, but the frequency $\eta$ of the jumps scales as $\eps\eta$, i.e. single jumps are rarer, thus less probable, in the longer time scale.

In order to get from~\eqref{eq:Boltzmann.g} an evolution equation in the $t$-scale, which avoids the detailed computation of the particle transitions in the unobservable $\tau$-scale, we introduce the scaled distribution function
$$ f(x,\,a,\,t):=g(x,\,a,\,t/\eps), $$
which satisfies the relations $\int_0^1\int_\Omega f(x,\,a,\,t)\,dx\,da=1$ and $\partial_t f=\frac{1}{\eps}\partial_\tau g$, and, by~\eqref{eq:Boltzmann.g},  the equation
\begin{align}
	\begin{aligned}[t]
		\frac{d}{dt} & \int_0^1\int_\Omega\varphi(x,\,a)f(x,\,a,\,t)\,dx\,da \\
		&= \frac{\nu}{\eps}\int_0^1\int_0^1\int_\Omega\int_\Omega(\varphi(x,\,a^\ast)-\varphi(x,\,a))
			h(x,\,y)f(x,\,a,\,t)f(y,\,b,\,t)\,dx\,dy\,da\,db \\
		&\phantom{=} +\frac{\mu}{\eps}\int_0^1\int_\Omega(\varphi(x,\,a^{\ast\ast})-\varphi(x,\,a))f(x,\,a,\,t)\,dx\,da \\
		&\phantom{=} +\eta\int_0^1\int_\Omega\int_0^{1-a}(\varphi(x,\,a^{\ast\ast\ast})-\varphi(x,\,a))
			P(t,\,j\vert x,\,a)f(x,\,a,\,t)\,dj\,dx\,da,
	\end{aligned}
	\label{eq:Boltzmann.f}
\end{align}
where we have defined $P(t,\,j\vert x,\,a):=p(t/\eps,\,j\vert x,\,a)$.

Taking $\varphi\in C^\infty(\Omega\times [0,\,1])$, with $\varphi(\cdot,\,0)=\varphi(\cdot,\,1)=0$, we expand
\begin{align*}
	\varphi(x,\,a^\ast)-\varphi(x,\,a) &= \eps\partial_a\varphi(x,\,a)\mathcal{G}(x,\,a,\,b)+
		\frac{\eps^2}{2}\partial^2_a\varphi(x,\,\tilde{a})\mathcal{G}^2(x,\,a,\,b), \\
	\varphi(x,\,a^{\ast\ast})-\varphi(x,\,a) &= \eps\partial_a\varphi(x,\,a)\mathcal{S}(x,\,a,\,u(x,\,\tau))+
		\frac{\eps^2}{2}\partial^2_a\varphi(x,\,\bar{a})\mathcal{S}^2(x,\,a,\,u(x,\,\tau))
\end{align*}
with $\tilde{a},\,\bar{a}\in (0,\,1)$, then we plug into~\eqref{eq:Boltzmann.g} to get
\begin{align*}
	\frac{d}{dt} & \int_0^1\int_\Omega\varphi(x,\,a)f(x,\,a,\,t)\,dx\,da \\
	&= \nu\int_0^1\int_0^1\int_\Omega\int_\Omega\partial_a\varphi(x,\,a)\mathcal{G}(x,\,a,\,b)h(x,\,y)
		f(x,\,a,\,t)f(y,\,b,\,t)\,dx\,dy\,da\,db \\
	&\phantom{=} +\mu\int_0^1\int_\Omega\partial_a\varphi(x,\,a)\mathcal{S}(x,\,a,\,u(x,\,\tau))f(x,\,a,\,t)\,dx\,da \\
	&\phantom{=} +\eta\int_0^1\int_\Omega\int_0^{1-a}(\varphi(x,\,a^{\ast\ast\ast})-\varphi(x,\,a))
		P(t,\,j\vert x,\,a)f(x,\,a,\,t)\,dj\,dx\,da \\
	&\phantom{=} +R(\eps),
\end{align*}
where the remainder $R$ is
\begin{align*}
	R(\eps) &:= \frac{\nu\eps}{2}\int_0^1\int_0^1\int_\Omega\int_\Omega\partial^2_a\varphi(x,\,\tilde{a})
		\mathcal{G}^2(x,\,a,\,b)h(x,\,y)f(x,\,a,\,t)f(y,\,b,\,t)\,dx\,dy\,da\,db \\
	&\phantom{:=} +\frac{\mu\eps}{2}\int_0^1\int_\Omega\partial^2_a\varphi(x,\,\bar{a})
		\mathcal{S}^2(x,\,a,\,u(x,\,\tau))f(x,\,a,\,t)\,dx\,da.
\end{align*}
Using that $0\leq h(x,\,y)\leq 1$ and $\int_0^1\int_\Omega f(x,\,a,\,t)\,dx\,da=1$, we see that
$$ \abs{R(\eps)}\leq \frac{\eps}{2}\norm{\partial_a^2\varphi}_\infty\left(\nu\norm{\mathcal{G}}_\infty^2+\mu\norm{\mathcal{S}}_\infty^2\right). $$
Therefore, if $\mathcal{G}$ and $\mathcal{S}$ are bounded,  $R(\eps)\to 0$ as $\eps\to 0^+$, and we obtain the equation
\begin{align}
	\begin{aligned}[t]
		\frac{d}{dt} & \int_0^1\int_\Omega\varphi(x,\,a)f(x,\,a,\,t)\,dx\,da \\
		&= \int_0^1\int_\Omega\partial_a\varphi(x,\,a)v[f,\,u](x,\,a)f(x,\,a,\,t)\,dx\,da \\
		&\phantom{=} +\eta\int_0^1\int_\Omega\int_0^{1-a}(\varphi(x,\,a^{\ast\ast\ast})-\varphi(x,\,a))
			P(t,\,j\vert x,\,a)f(x,\,a,\,t)\,dj\,dx\,da
	\end{aligned}
	\label{eq:hydro.1}
\end{align}
for
\begin{equation}
	v[f,\,u](x,\,a):=\nu\int_0^1\int_\Omega \mathcal{G}(x,\,a,\,b) \, h(x,\,y)
		f(y,\,b,\,t)\,dy\,db+\mu\mathcal{S}(x,\,a,\,u(x,\,\tau)).
	\label{eq:velocity}
\end{equation}

Let us further inspect the last term at the right-hand side of~\eqref{eq:hydro.1}. Substituting $j$ with $a^{\ast\ast\ast}$ according to~\eqref{eq:int.rules} yields:
\begin{align*}
	\int_0^1 & \int_\Omega\int_0^{1-a}\varphi(x,\,a^{\ast\ast\ast})P(t,\,j\vert x,\,a)f(x,\,a,\,t)\,dj\,dx\,da \\
	&= \int_0^1\int_\Omega\int_a^1\varphi(x,\,a^{\ast\ast\ast})P(t,\,a^{\ast\ast\ast}-a\vert x,\,a)f(x,\,a,\,t)\,da^{\ast\ast\ast}\,dx\,da \\
\intertext{whence, switching the integrals in $a^{\ast\ast\ast}$ and $a$,}
	&= \int_0^1\int_\Omega\varphi(x,\,a^{\ast\ast\ast})\left(\int_0^{a^{\ast\ast\ast}}P(t,\,a^{\ast\ast\ast}-a\vert x,\,a)
		f(x,\,a,\,t)\,da\right)\,dx\,da^{\ast\ast\ast}.
\end{align*}

On the whole~\eqref{eq:hydro.1} rewrites as
\begin{align*}
	\frac{d}{dt} & \int_0^1\int_\Omega\varphi(x,\,a)f(x,\,a,\,t)\,dx\,da \\
	&= \int_0^1\int_\Omega\partial_a\varphi(x,\,a)v[f,\,u](x,\,a)f(x,\,a,\,t)\,dx\,da \\
	&\phantom{=} +\eta\int_0^1\int_\Omega\varphi(x,\,a)\left(\int_0^aP(t,\,a-a_\ast\vert x,\,a_\ast)f(x,\,a_\ast,\,t)\,da_\ast
		-f(x,\,a,\,t)\right)\,dx\,da,
\end{align*}
which can be finally recognised as a weak form of
$$ \partial_t f+\partial_a(fv[f,\,u])=J[f]. $$

This equation is a balance law with non-local transport velocity given by~\eqref{eq:velocity}. The term $J[f]$ at the right-hand side is the \emph{jump operator}
$$ J[f](t,\,x,\,a):=\eta\left(\int_0^aP(t,\,a-a_\ast\vert x,\,a_\ast)f(\,x,\,a_\ast,\,t)\,da_\ast-f(x,\,a,\,t)\right), $$
which, owing to~\eqref{eq:int.p}, is such that $\int_0^1\int_\Omega J[f](t,\,x,\,a)\,dx\,da=0$. Denoting the probability law of the jumps as $P(t,\,a_\ast\to a\vert x)$ we see that it is the term modelled in~\eqref{J formula}, with the dependence on the spatial distribution
of the jumps hidden in the dependence of $P$ on $x$.

\begin{remark}
If we assume that the neural network is composed by an extremely large number of neurons, which are connected mainly with other close neurons,  we can take the probability $h$ as $h(x,\,y)=\chi_{B_{1/N}(x)}(y)$, where $N$ is the total number of neurons and $B_{1/N}(x)\subset\Omega\subset\rn$ is the ball centred in the point $x$ with radius $1/N$. If we suppose also $\nu\propto N^d$ (i.e. the more the neurons the more frequent the prionic transmission of the disease among them) then in the limit $N\to\infty$ we obtain that $\nu h(x,\,y)\to\delta_0(y-x)$ and 
$$ v[f,\,u](x,\,a)=\int_0^1\mathcal{G}(x,\,a,\,b)f(x,\,b,\,t)\,db+\mu\mathcal{S}(x,\,a,\,u(x,\,\tau)), $$
which is indeed the form of $v$ postulated in~\eqref{velocity bis}.

Notice that, in general, $v$ depends on contributions from multiple time scales, in fact the prionic transmission of the disease (first term at the right-hand side of the formula above) takes place in the slower $t$-scale while the poisoning of the neurons by A$\beta$ polymers (second term at the right-hand side of the formula above) takes place in the faster $\tau$-scale.
\end{remark}

\bigskip

\section{Discussion}\label{discussion} 

There are two main features of our macroscopic model that deserve to be highlighted: first of all, as we showed above, it derives mathematically from a microscopic model we carefully developed relying on an accurately selected set of phenomena described in biomedical literature. This has been possible also thanks to the interdisciplinary character of
our team, including mathematicians and medical doctors. 
Secondly,  the model is characterised by a high level of flexibility, which potentially allows one to simulate different 
working hypotheses and compare them with clinical data. In fact, the model provides a flexible tool to test alternative conjectures on the evolution of the disease. 
Up to now we have chosen  some specific aspects of the illness, such as aggregation, diffusion and removal of $A\beta$,  possible spread of neuronal damage in the neural pathway, and, to describe the onset of AD, a random neural deterioration mechanism. Some mathematical results have been obtained, and numerical simulations have been compared with clinical data. Although we have restricted ourselves to a 2-dimensional rectangular section of the brain, the results are in good qualitative agreement with the spread of the illness in the brain at various stages of its evolution. In particular, the model  captures the cerebral damage in the early stage of MCI. 

There are several research issues that we would like to address in the future. As mentioned in the Introduction, 
the main shortage of the present model is the complete omission of the action of the $\tau$ protein and the microglia
 (in this context we mention a mathematical model proposed in \cite{HHPW}).
The mechanisms related to the presence of the two substances should eventually be considered in a subsequent 
evolution of this model, both to obtain optimal quantitative agreement with clinical data
and also to investigate the possible formation of patterns in the distribution of the level of malfunctioning of the brain.

From a numerical point of view, simulations should become more realistic, in a 3-dimensional domain which matches the geometric characteristics of the brain.
Moreover, a certain sensitivity of the numerical output to the value of the constant $\gamma$ in~\eqref{eq:bc_Robin}, which models the removal of $A\beta$ through the choroid plexus, spontaneously leads to the question whether dialysis-mechanisms can be introduced to enhance $A\beta$-removal artificially. Most probably, a serious answer to this question requires, in addition to a detailed comparison with clinical data, a more refined 
 modelling 
of the clearance of soluble A$\beta$ by the cerebral fluid.

In \cite{AFMT} and \cite{BFMTT_MMB} the reader can find an exhaustive discussion of the
literature on mathematical models for AD.
Besides that, in the recent paper \cite{HF} a large system of reaction-diffusion equations is proposed as a macroscopic model   
which takes into account many of the processes which possibly play a role in the development 
of AD. In addition the paper contains some simulations of medical treatments. 
The authors do not consider the 
onset of AD, and model the action of the $\beta$-amyloid in a way which is quite different from our approach. 
In a forthcoming paper we shall compare both approaches.
\medskip

\section*{Acknowledgments}
This paper is dedicated to Stu with deep affection and gratitude.

The authors would like to express their gratitude to MD Norina Marcello for many stimulating and fruitful discussions over several years.

B.~F. and M.~C.~T. are supported by the University of Bologna, funds for selected research topics, and by MAnET Marie Curie
Initial Training Network. B.~F. is supported by 
GNAMPA of INdAM (Istituto Nazionale di Alta Matematica ``F. Severi''), Italy, and by PRIN of the MIUR, Italy.

A.~T. is member of GNFM (Gruppo Nazionale per la Fisica Matematica) of INdAM (Istituto Nazionale di Alta Matematica ``F. Severi''), Italy.

\bibliographystyle{plain}
\bibliography{references}
\end{document}